\documentclass[letterpaper]{article} % DO NOT CHANGE THIS
\usepackage{aaai2026}  % DO NOT CHANGE THIS
\usepackage{times}  % DO NOT CHANGE THIS
\usepackage{helvet}  % DO NOT CHANGE THIS
\usepackage{courier}  % DO NOT CHANGE THIS
\usepackage[hyphens]{url}  % DO NOT CHANGE THIS
\usepackage{graphicx} % DO NOT CHANGE THIS
\usepackage{natbib}  % DO NOT CHANGE THIS AND DO NOT ADD ANY OPTIONS TO IT
\usepackage{caption} % DO NOT CHANGE THIS AND DO NOT ADD ANY OPTIONS TO IT
\usepackage{algorithm}
\usepackage{algorithmic}

\usepackage{newfloat}
\usepackage{listings}
\DeclareCaptionStyle{ruled}{labelfont=normalfont,labelsep=colon,strut=off} % DO NOT CHANGE THIS
\floatstyle{ruled}
\newfloat{listing}{tb}{lst}{}
\floatname{listing}{Listing}
\title{AI-based Traffic Modeling for Network Security and Privacy: Challenges Ahead}
\author{
 Dinil Mon Divakaran
}
\affiliations{
    A*STAR Institute for Infocomm Research (A*STAR I$^2$R) \\
    dinil\_divakaran@a-star.edu.sg
}

\usepackage{xcolor}
\usepackage{xspace}

\usepackage[most]{tcolorbox}

\newcommand{\netsec}{{\texttt{NetS\&P}}\xspace}

\newtcolorbox{cooltextbox}[1][]{%
    colback=blue!0,
    colframe=blue!0,
    notitle,
    sharp corners,  
    borderline west={3pt}{0pt}{blue!70!red},
    enhanced,
    breakable,
    left=3pt,
    right=0pt,
    top=-0.75pt,
    bottom=0pt
    }

\begin{document}

\maketitle

\begin{abstract}
Network traffic analysis using AI (machine learning and deep learning) models made significant progress over the past decades. Traffic analysis addresses various challenging problems in network security, ranging from detection of anomalies and attacks to countering of Internet censorship. AI models are also developed to expose user privacy risks as demonstrated by the  research works on fingerprinting of user-visiting websites, IoT devices, and different applications, even when payloads are encrypted. 

Despite these advancements, significant challenges remain in the domain of network traffic analysis to effectively secure our networks from evolving threats and attacks. After briefly reviewing the relevant tasks and recent AI models for traffic analysis, we discuss the challenges that lie ahead. 
\end{abstract}

% Uncomment the following to link to your code, datasets, an extended version or similar.
% You must keep this block between (not within) the abstract and the main body of the paper.
% \begin{links}
%     \link{Code}{https://aaai.org/example/code}
%     \link{Datasets}{https://aaai.org/example/datasets}
%     \link{Extended version}{https://aaai.org/example/extended-version}
% \end{links}

\section{Introduction}
% no \IEEEPARstart

%Networks form the backbone of modern communications and must be protected from ever-evolving cyber threats and attacks. 
The sophistication of networks attacks has increased over time due to the rapid proliferation of new technologies, applications, network protocols and devices. Meanwhile, network traffic keeps increasing, both in volume and speed, further complicating security challenges. To effectively secure the networks, traffic analysis has long been recognized as a critical component. 
As payloads in network packets are hardly available for signature-based detection with the high adoption of TLS by different applications~\cite{Google-transparency-mail, Google-transparency-HTTP}, statistical models are becoming inevitable to analyze network traffic. 
Decades of research efforts have led to the development of various statistical, machine learning (ML) and deep learning (DL) models of different capabilities to tackle multiple  network security and privacy (\netsec) tasks. % (Section~\ref{sec:tasks-models}).  

DL models, in particular, are valuable for network traffic analysis in \netsec tasks for two key reasons, or rather, to address two significant challenges.
First, network data is enormous; for example, a 1~Gbps link generates hundreds of gigabytes of data per day under high utilization, even when considering only packet headers (excluding the payload). And network bandwidth has increased steadily, with even consumer (broadband) bandwidth now reaching 1~Gbps. Meanwhile, enterprises and telcos/ISPs utilize bandwidth capacities that are orders of magnitude higher. Second, the number of features that can be extracted from network traces can potentially reach several hundreds or even thousands in count. For instance, when modeling a user browsing session as a data point for training/inference, a small session can easily consist of 200 packets, with each packet containing 10 attributes (e.g., packet size, inter-arrival time, and so on). Consequently, representing this relatively short session requires 2,000 features. As we know, DL models are able to process and learn useful patterns from huge datasets with large feature space.

In light of the current state-of-the-art of AI in network traffic modeling, this work discusses the key challenges that lie ahead in securing networks and protecting user privacy. 

\section{\netsec: Tasks and Models}
\label{sec:tasks-models}

We discuss important \netsec tasks and AI-based solutions in the literature, without aiming for exhaustiveness.

\subsection{Anomaly detection} 

Anomalous behaviors occurring on endpoints that manifest in network traffic are the  anomalies we aim to detect. In an enterprise network, such anomalies may arise from the adoption of new applications (e.g., ChatGPT), sudden use of new protocols---say, encrypted DNS protocols such as DNS-over-HTTPS (DoH), 
%~\cite{doh} 
or DNS-over-TLS (DoT),
%~\cite{dot}, 
introduction of new devices (e.g., smartwatches), and so on. At homes or hotel rooms, the presence of hidden IoT devices is a security anomaly~\cite{Lumos-hidden-IoT-USENIX-SEC-2022}. Malicious activities such as bot communications, malware traffic and DDoS attacks are also anomalies. For network administrators, these anomalies pose potential threats that must be detected and analyzed to determine appropriate mitigation actions, such as implementing new policies for newly introduced devices or blocking IP addresses associated with identified bots. 

\noindent{\bf AI models}: The primary challenge for this task stems from the diverse types of network anomalies, including those that are previously unknown.  
For instance, while the recent encrypted DNS protocols like DoH provide better privacy for users, they also offer malware developers new ways to hide their communications~\cite{survey-encrypted-DNS-mw-2022}, say, for exfiltrating sensitive information~\cite{real-time-DNS-exfiltration-NDSS-2024}. These are relatively new anomalies unknown before DoH was deployed.
Existing research in literature tackles this challenge of anomaly detection by developing unsupervised and semi-supervised models.
Early research works explored PCA (Principal Component Analysis),  statistical hypothesis testing and regression models to detect anomalous patterns in network traffic~\cite{PCA-Diot-2004, brauckhoff2009applying, NADA-2018, evidence-gathering-2017}. With the emergence of generative deep learning models for unsupervised problem settings~\cite{AD-robust-AE-2017, adversarially-learned-AD-ICDM-2018}, we witnessed the development of reconstruction-based unsupervised network anomaly detection solutions in the last decade~\cite{KitSUne-2018-NDSS, GEE-2019}. For example, GEE~\cite{GEE-2019} trains a variational autoencoder (VAE) on noisy benign network traffic sessions to learn its corresponding representation in the latent space, which it subsequently uses to detect deviating behaviors during the inference stage. 

\subsection{Attack classification} While anomaly detection solutions are useful for detecting broad and unknown types of threats and attacks, they do not identify the specific attack type, which is important to decide on the appropriate mitigation strategy. For example, DDoS traffic must be immediately blocked as close to the sources as possible (a challenge in itself); whereas if a malware is detected, the infected machine has to be contained, incident response initiated, and forensic analysis triggered. Identifying the type of attack is achieved by classifying the network data into one or more {\em known} attack classes of interest, such as botnet activity, DDoS attacks, C\&C communications, and password spraying, among others. Anomaly detection and attack classification typically complement each other, e.g., as demonstrated in~\cite{ADEPT-2021}, enabling i)~the detection of both known and unknown attacks, and ii)~the identification of attack types. 

\noindent{\bf AI models}: %As long as such labeled datasets are available, the models (based on the well-known features) are known to perform with (or, have reported) high accuracy. 
While the problem might appear to be a multi-class classification problem, the data available for different attacks vary. DDoS requires simple features that capture the traffic rate, whereas detecting the presence of malware that uses DGA (domain generation algorithm) requires analysis of (plain-text) DNS payload.
Therefore, solutions in this space typically 
train binary classifiers to detect specific attacks. For example, several ML algorithms for DDoS detection are evaluated considering both accuracy and inference latency in~\cite{DDoS-eval-latency-ICNP-2024}. Similarly, prior works have proposed methods for classifying DGA domains~\cite{DGA-Rossow-2024}, identifying bot traffic to servers using GAN-based data augmentation~\cite{botnet-S&P-2020}, and detecting encrypted malware communications over TLS~\cite{anderson2017machine} as well as Tor~\cite{Tor-malware-CCS-2022}.

\subsection{IoT device identification} Identifying IoT devices connected to a network is necessary for security auditing and  for configuring appropriate policies  (e.g., firewall rules) for effective management of the network. The challenge lies in determining the type of devices connected to the network through {\em traffic analysis}. In an enterprise network, a device identification solution would analyze north-south  traffic as well as east-west  traffic to identify devices that communicate both internally and externally.  When labels for IoT devices and their corresponding network data are available, the task transforms into a multi-class classification problem. % from a modeling perspective. \\

\noindent{\bf AI models}: IoT device identification has received considerable attention over the past decade, resulting in the proposal of various supervised and semi-supervised approaches for classifying known and unknown devices, respectively. They include several works that are based on conventional ML as well as DL models~\cite{2017-profileiot, 2019-DIoT, DEFT-2019, IoT-biLSTM-2020, WU-ZEST}. A recent work~\cite{usenix-2020-youarewhatyou} demonstrates that devices connected to a WiFi network can be identified by training a DL model on the broadcast/multicast traffic observed. 
%\footnote{While the aforementioned works (and much of the research discussed in this paper) focus on IP-level traffic modeling, it is worth noting that there are also proposals for fingerprinting devices using RF (radio frequency) signals across various communication technologies (e.g., LoRa, ZigBee, etc.) using different neural network architectures~\cite{shen2021radioFP, RF-LoRa-LSTM-2021, cross-domain-RF-fingerprinting-INFOCOM-2024}.}.

Device identification is considered a defensive security measure, as organizations~\cite{WU-ZEST} and individuals~\cite{Lumos-hidden-IoT-USENIX-SEC-2022} need to track the devices connected to their networks.  However, when viewed from the perspective of a home or public user, adversarial identification of consumer devices---whether using a compromised router or through a public WiFi hotspot---represents a violation of user privacy. This has led to the development of counter-fingerprinting techniques, e.g., using generative adversarial perturbations%~\cite{poursaeed2018generative}, to defeat privacy attacks
~\cite{shenoi2023ipet}.

%In general, we observe that the modeling aspects (features defined and models used) for this task significantly overlap with those proposed for anomaly detection and attack classification. A distinction is that the solutions developed for IoT device identification are specifically trained and tested on data generated by broader class of IoT devices, such as cameras, mobile devices, sensors, switches, laptops, etc. (e.g., see~\cite{2019sosrIoT}).

\subsection{Website fingerprinting (WFP) attack} WFP attack targets user privacy by aiming to identify the websites a user visits. The research efforts in this space shed light on the vulnerability of existing and new network protocols in revealing sensitive user information. Since the focus is on web traffic, the relevant network protocols for traffic modeling are primarily limited to DNS (and its encrypted variants) and HTTPS. However, as users may employ VPNs or Tor~\cite{dingledine2004tor} to mitigate such privacy attacks, research proposals typically assume that the traffic is tunneled through these services. The targets of WFP attacks can include anyone, such as citizens of interest, journalists, activists, or politicians, who may be monitored (or surveilled) by those in power. Consequently, the threat model assumes an adversary that has access to a network node, such as a router (either through direct control or compromise). Thus the adversary can passively monitor and analyze network traffic to identify the websites visited by the target users.

\noindent{\bf AI models}: Given that a browsing session consists of hundreds or even thousands of packets, this topic has witnessed significant development of deep learning-based attack strategies. In other words, the ability of deep learning models to learn from a large feature space represents a substantial advantage for this task. 
The literature has interesting WFP attack proposals that use various DL architectures ranging from CNN to AE to sequence models such as LSTM and transformer~\cite{NDSS-2018-wfp-lstm, pfp-PETS-2019, varCNN-WFP-2019, Usenix-2022-onlinewebsitefingerprinting, webpage_quic, multi-tab-FP-transformer-SP2023, transformer-WFP-CCS-2023, MIL-WFP-USENIX-2024, Levi-2024}. There are also strategies proposed for countering them~\cite{gong2020FRONTGLUE, smith2022QCSD, GAN-WFP-defense-SP-2022, siby2023PET, Usenix-2023-website-fingerprinting}. These models are evaluated under two scenarios: i)~a {\em closed-world} 
 (lab) setting, where a classification model is trained to classify $n$ specific {\em monitored} websites (each with samples collected over time) and is tested for accurate classification of the same $n$ websites, where $n$ is typically in a few hundreds; ii)~a more realistic {\em open-world} setting, where the challenge is to identify $n$ {\em monitored} websites under the assumption that the user visits $m~(m\gg n)$ {\em unmonitored} websites, where $m$ is in tens of thousands. Since tunneling of traffic hides packet headers, (different from the previous tasks) the features used here are the basic (or {\em raw}) ones such as packet-size, inter-arrival time (IAT) (or time lag) of packet  and direction of packet (as browsing generates bidirectional traffic). 
% It is worth noting that similar approaches are also taken in analysis of tunneled traffic for security purposes, such as detection of malware that use Tor for communications~\cite{Tor-malware-CCS-2022}, detection of tunneled flooding traffic~\cite{tunneled-flooding-CCS-2024}, etc.

\subsection{Other tasks and models}

There are a numerous other \netsec tasks; due to space constraints, we provide a brief outline here. 

\begin{itemize}

    \item {\bf Censorship and Anonymity.}
    Nation-state adversaries carry out traffic analysis for surveillance and censorship. The Tor anonymity network~\cite{dingledine2004tor} is developed to offer anonymized access to Internet users. However, Tor is prone to de-anonymization attack~\cite{nasr2018deepcorr, oh2022deepcoffea}. 
    %In one of the recent attacks~\cite{oh2022deepcoffea}, a triplet network (rather, a tripplet loss function) is utilized to correlate flows at the entry and exit nodes of a Tor network, enabling linkability to users. 
    A most recent proposal~\cite{RECTor-2025} slices flows into windows and employs attention-based MIL (multi-instance learning) to learn the most relevant segments of potentiality noise traffic flows, so as to correlated flows at the entry and exit nodes of a Tor network.
    %Models are also developed for countering censorship as well as to understand censorship strategies~\cite{detecting-proxies-NDSS-2020}. 

    \begin{comment}
        
    \item {\bf (IP) Network traffic fingerprinting.} WFP (discussed above) can be seen as one specific attack under the broader class of network fingerprinting attacks (over IP networks since features extracted, e.g., packet size, are similar at IP layer). Examples of such attacks include video fingerprinting~\cite{schuster2017beauty, sabzi2024netshaper, gast2024snailload}, application fingerprinting~\cite{FS-Net-traffic-classifier-INFOCOM-2019, unknown-app-classification-2020, dapp-GNN-FP-2021, traffic-classification-KDD-2023}, etc. 
    Application fingerprinting itself covers a few problems; for example, the proposal in~\cite{dapp-GNN-FP-2021} employs a GNN  (Graph Neural Network) to model and
    fingerprint dApps (decentralized applications) on blockchain platforms. There are also DL-bases techniques proposed to classify applications (e.g., Skype, Youtube, etc.) based on the network traffic~\cite{FS-Net-traffic-classifier-INFOCOM-2019, unknown-app-classification-2020}---such solutions serve as an attack on user privacy, although they are also considered useful for differentiated QoS provisioning. 
    \end{comment}
    
    \item {\bf Token inference attack.} In this attack, the encrypted traffic is passively monitored to estimate the token sizes from response packets, to subsequently use to to infer the response from an AI assistant~\cite{prompt-infer-USENIX-SEC-2024}, using a fine-tuned T5 transformer model~\cite{T5} is fine-tuned. 
    %for the token inference attack in~\cite{prompt-infer-USENIX-SEC-2024}.

\begin{comment}

    \item {\bf Privacy attacks on cellular traffic.} DL models are also developed for privacy attack on users of cellular traffic. For example, in~\cite{cellular-video-2022}, the authors demonstrate a video fingerprinting attack on UE (user equipment) in LTE network using multiple CNN classifiers trained on passively monitored (layer~2) radio traffic; similarly website fingerprinting attack using multiple ML and DL models are shown to be effective on users in LTE network~\cite{kohls2019lost}. 
        
\end{comment}
\end{itemize}

We also highlight the application of adversarial ML, which aims to generate adversarial samples for evading systems utilizing AI models. In the context of \netsec, an adversarial ML approach can be exploited not only to bypass AI-based security systems but also to counter threats related to privacy and censorship. For instance, authors in~\cite{shenoi2023ipet} adapt generative adversarial perturbations~\cite{poursaeed2018generative} for evading IoT fingerprinting solutions that employ AI models. And in the space of website fingerprinting, a defense based on GAN, specifically, WGAN-div~\cite{wu2018wasserstein}, to evade WFP attacker (discriminator) is studied in~\cite{GAN-WFP-defense-SP-2022}. 
%GAN-based approach is also proposed to transmit sensitive data over multimedia protocol (for audio) to counter censorship~\cite{jia2023voiceover-FOCI}. 

\section{Challenges (and Opportunities) Ahead}
\label{sec:challenges}

\subsection{Data challenges}

One of the most important questions is, {\em do the current AI-based solutions generalize beyond the experiments conducted in controlled lab environments?}
Addressing this question is challenging due to the lack of {\em high-quality} {\em labeled} data. Most research on network anomaly detection, attack classification, and device identification relies on openly available labeled datasets. However, since these datasets are generated in lab environments, they often contain artifacts or `bad design smells'~\cite{bad-smells-EuroSP-2024} that raise concerns on model over-fitting and biases. 
In ~\cite{bad-smells-EuroSP-2024}, the authors analyze seven highly-cited datasets 
and provide evidence of such artifacts in network intrusion detection research. For instance, in one dataset with eight attack classes, two basic features exhibit minimal overlap between benign and attack traffic. As a result, a simple perturbation of these two features---unrelated to the inherent characteristics of attacks---can evade multiple ML and DL models. This simple approach outperforms a state-of-the-art adversarial attack~\cite{adversarial-NIDS-2022} employing evolutionary computation and generative adversarial networks. This finding suggests that evaluations in~\cite{adversarial-NIDS-2022} may not provide a ``meaningful measure of the attack's effectiveness''~\cite{bad-smells-EuroSP-2024} due to dataset flaws. Similar biases in network datasets have also been highlighted for other \netsec tasks, e.g., website fingerprinting~\cite{data-explainable-WFP-PETS-2023, retrace-Tor-2024, genuineTorTraces-2024}.

The issue at hand is not about the existing proposals per se, 
but rather the quality of the datasets used for evaluating network security tasks.
Going one step further, the fundamental challenge is the difficulty of validating the quality network traffic.  The traffic generated by a single day's communications in an enterprise network can reach hundreds of gigabytes, and when considering this scale over multiple days, across various enterprises and different verticals, the volume becomes staggering. Therefore, collecting and labeling real-life network traffic flows is an obstacle. Furthermore, real-world network traffic contains sensitive user information, and sharing this data poses significant privacy risks and ethical concerns. Model-based attacks can extract hidden patterns as fingerprinting attacks  in the previous section reveal; meta-data from packet headers (excluding payloads) is sufficient to infer sensitive information. Consequently, it is understandable that enterprises and telco operators refrain from publicly sharing traffic collected from their networks for research and development purposes.

\begin{cooltextbox}
\textrm{There is currently a lack of real-world quality datasets with labels for \netsec tasks such as anomaly detection, attack detection, website fingerprinting, etc.}
\end{cooltextbox}

This leaves the research community with the second-best option---generating traffic datasets in controlled lab environments.
There are promising directions here. The aforementioned in-depth analysis of open datasets represents an important step forward; it also sheds light on how to evaluate the quality and fidelity of lab-generated datasets. Learning of broad principles will guide in framing the right questions for specific \netsec tasks. For example, if SSH brute-force attempts are present in the attack category, the benign category should also have normal SSH flows; otherwise encoding the port (22) would classify all SSH flows as malicious and (falsely) validate the model! 

\begin{cooltextbox}
\textrm{We need to establish design principles for generating real traffic data in controlled environments, and the data generated should be evaluated for fidelity. }
\end{cooltextbox}

As second direction, we note several recent research proposals put forth to {\em synthesize} traffic data with high fidelity~\cite{botnet-S&P-2020, GAN-pkt-synthesis-SIGCOMM-2022, data-synthesis-DCN-HotNets-2023, Zoom2Net-SIGCOMM-2024, jiang2024netdiffusion, flowChronicle-2024}. In~\cite{data-synthesis-DCN-HotNets-2023}, the authors propose using a Variational Autoencoder (VAE) to learn the flow-size distribution from datacenter network traces, subsequently generating sequences of flow sizes with a Recurrent Neural Network (RNN) using Gated Recurrent Units (GRU). In~\cite{Zoom2Net-SIGCOMM-2024}, a transformer model is used to estimate missing data in a network telemetry time-series data by correlating information from multiple sources. Although they do not directly address security problems, we can take a leaf out of them to understand the key requirements that have to be met to synthesize data for \netsec tasks:
\begin{cooltextbox}
%{Research questions for traffic synthesis:}
Synthetic Data Generation:
\begin{itemize}
    \item How critical is the leakage of distribution parameters in the context of enterprise network data? 
    \item How can models learn with minimal linkage of distribution parameters to the source network? 
    
    \item What impact does constrained data synthesis have on performance of \netsec solutions?
    
\end{itemize}
\end{cooltextbox}

Another promising direction is {\em emulating} applications that generate network traffic in a controlled environment. This is a common practice in website fingerprinting domain, where thousands of websites are visited (over Tor and VPN) using browsers, to generate the corresponding browsing traffic (e.g., HTTPS packets). However, for training models for securing a consumer or enterprise network, broader set of applications have to be emulated. A recent proposal explores the possibility of generating network traffic by orchestrating public GitHub repositories~\cite{gen-nw-traffic-2022}. Building on this concept, the authors of ~\cite{khan2024harnessing}  propose a data-generation platform that can be managed using SDN. 
%This new direction is promising for addressing the challenges associated with generating realistic data. 
We are still in early stages, and further research is needed to validate the traffic thus generated, and subsequently also to explore  generating malware and anomalous traffic with high fidelity and diversity.

\begin{cooltextbox}
%{Research questions for traffic synthesis:}

Emulation of benign and malicious traffic: 
\begin{itemize}

    \item How do we emulate applications corresponding to different enterprises (e.g., finance vs. academia) with varying user characteristics, for generating benign application data?  
    
    \item How do we carefully and ethically emulate malware behavior, including infection, lateral movement, C\&C communications, data exfiltration, and other activities, for the purpose of generating malicious data? 
    
    \item How can SDN help to dynamically configure and manage networks for traffic generation?~\cite{netreplica-2025}
    
\end{itemize}
\end{cooltextbox}

\subsection{Practical deployment}

Real-time per-packet inference in network traffic analysis faces several practical limitations. Enterprise networks operate at tens of Gbps whereas telco networks operate at 100s of Gpbs. Even at 10~Gbps, the time available for per-packet decision-making is constrained to less than 100 nanoseconds per packet, which poses challenges for complex computations required for DL-based solutions. In fact, an expensive step in building an ML/DL pipeline is parsing packets for high-speed feature extraction~\cite{Unsenix2023-IDP-programmable-switch-flow}.

A single packet typically does not provide sufficient information to make informed decisions about its legitimacy. For instance, a TCP SYN packet can be part of a legitimate connection or a malicious attempt.Only by {\em aggregating} packets, e.g., by source/destination IP address/port and segmented by time, can patterns indicative of attacks, such as TCP SYN floods, be identified. However, storing and processing these {\em packet aggregates} require substantial resources, a challenge that increases with the network rate.

\begin{cooltextbox}
Packet aggregates serve as meaningful units for modeling traffic. However, extracting information---such as total volume or duration---from packet aggregates (e.g., 5-tuple flows or sessions grouped by src/dst IP) still necessitates processing each individual packet, which becomes increasingly challenging as network rates rise.
\end{cooltextbox}

In this context, programmable data planes offer promising solutions for in-network computation at rates of terabits per second (Tpbs)~\cite{in-network-dumb-switch-hotnets-2017, zhang2020poseidon, Usenix-DDos-programme-switch-2021, p4-survey-2021}. These are essentially switches and smartNICs that allow programmability of packet processing pipelines directly on the network devices while constraining to specific packet processing behavior, e.g., in terms of latency and throughput. Indeed, there have been attempts to implement tree-based models in programmable switches as that aligns well with the match-action pipeline of the programmable switch architecture~\cite{2019-do-switchdream, Unsenix2023-IDP-programmable-switch-flow, offloadingML-ACM-Surveys-2023, hybrid-inetwork-classification-2024}. 
%The question whether more sophisticated models, in particular deep learning models, can be implemented as in-network services~\cite{in-net-neural-2021} does have hope, given the advancement of programmable switches and languages~\cite{johnson2024sequence}, the availability of high-speed (programmable) SmartNICs~\cite{firestone2018azure, michalowicz2023battle} with hardware accelerators and early demonstration of capability to run binary neural networks~\cite{binary-NN-SmartNIC-2022}, as well as the continuous progress in developing packet processing techniques that scale to hundreds of Gbps~\cite{stateful-pkt-processing-NSDI-2023, stateless-pkt-header-ICNP-2024}. 

However, the cost of processing every single packet, even when decisions are based on packet aggregates, remains a significant challenge~\cite{Marina-ML-monitoring-Terabit-2024}. Sampling has long been a viable solution for various network tasks, such as heavy-hitter monitoring~\cite{sampling-elephants-2003}. Nonetheless, applying sampling to model network traffic for security tasks requires further investigation. 
Developing models that can effectively learn from and infer dynamic features is important. 
Additionally, we have to design intelligent sampling strategies tailored to security use cases. For instance, protocol handshakes (TCP, TLS, etc.) are carried out in the initial part of a connection, and therefore might reveal important information~\cite{C2-TLS-RAID-2024}. This raises a research question---would an {\em adapting} sampling strategy that samples different parts of a packet sequence at different probabilities be better than  static sampling methods~\cite{IPFIX}? 
%(such as sampling one packet out of every 100). T
To answer this, we need to explore the trade-offs between sampling methods and their impact on model performance in detecting security threats. With programmable data planes, we now have the possibility to develop an in-network solution that makes both decisions---sampling and inference---without leaving the data plane. 

\begin{cooltextbox}
    \begin{itemize}
        \item Programmable data planes hold promise for implementing in-network DL-based solutions.

        \item Intelligent sampling, combined with models that can effectively handle missing data points, need to explored for \netsec tasks.
        
    \end{itemize}
\end{cooltextbox}

\subsection{Explainability of predictions}

Explainability technically allows us to interpret the behavior of models, or in other words, decisions made by models. The capability to explain the model predictions is important in cybersecurity to build trust between model developer, end user and stakeholders~\cite{XAI-depoloyment-2020}. Explainable models also aid in improving performance by identifying and mitigating potential pitfalls, such as biases, vulnerabilities to out-of-sample data, and overfitting~\cite{Trustee-CCS-2022, bad-smells-EuroSP-2024}. 

Consider different cybersecurity use cases.  For instance, in malware analysis, an explainable model highlights specific API calls that perform malicious actions or endpoint registers that were modified, providing actionable insights for mitigation~\cite{XAI-malware-2024}. In phishing detection, an explanation-based solution can reveal discrepancies between the brand displayed on a webpage and its domain name, helping users understand why a website is flagged as suspicious~\cite{lee-LLM-phishing-2024}. For provenance graph modeling of endpoint events~\cite{prov-ninja-2023}, a SOC (Security Operations Center) analyst would find it helpful  if a model identifies the sequence of processes and events that led to the alert raised, such as {\em executing a dropped binary that connects to a malicious IP}.

The above examples illustrate the utility of {\em local explanations} that explain (the features leading to) inference decision for the given input samples. 
On the other hand, with {\em global explanations}, the goal is to decipher  the overall understanding of the model; e.g., knowing certain expensive features do not increase the detection capability is useful to reduce the cost of deployment~\cite{cost-aware-FS-IoT-2021, Marina-ML-monitoring-Terabit-2024}. We refer readers to~\cite{zhao2024explainability} for a detailed breakdown of these two categories of explanations and understanding of the current state-of-the-art techniques for explainability. Some of these techniques have been recently explored in the network security community, concentrating mostly on anomaly and attack detection~\cite{GEE-2019, Trustee-CCS-2022, xNIDS-USENIX-SEC-2023, global-explanations-AD-CCS-2024}. 
The state-of-the-art DL solutions for \netsec tasks, e.g.,~\cite{multi-tab-FP-transformer-SP2023, WU-ZEST, prompt-infer-USENIX-SEC-2024, RECTor-2025}, are primarily based on the transformer model architecture~\cite{vaswani2017attention}; besides, transformer-based foundation models are also emerging for \netsec tasks (see below). This indicates that the existing techniques and challenges related to explainability in transformer models are pertinent to \netsec. 
%~\cite{zhao2024explainability}.  

We also have to take consider the perspective of the end-user to understand what needs to be `explained' along with the inference results. And this is task dependent. 
%In website fingerprinting (or broadly network traffic fingerprinting), our goal is to understand the privacy risks to users, even when they use encrypted protocols (DoH and HTTPS) and tuneling solutions (VPN and ToR). In this case, it is useful to identify the features that allow a website (or video, app or device) to be recognized, enabling the research and development of counter-strategies~\cite{GAN-WFP-defense-SP-2022, shenoi2023ipet, sabzi2024netshaper}.
In anomaly or attack detection, a SOC  analyst needs to understand why a particular traffic session is deemed malicious. This requires going beyond ranking important features (used for input representation)~\cite{GEE-2019, Trustee-CCS-2022}. 
Indeed the prediction should be accompanied with concrete explanations, such as noting that `there is a series of connection attempts from this enterprise endpoint that failed,' indicating a C\&C connection attempt from an infected endpoint. A potential approach might be to utilize LLMs to `translate' input feature-level explanations to a format consumable by SOC analysts. 

Explainability also provides insights into adversarial ML attack capability (and thereby defense); e.g., DGA classifiers are known to be prone to adversarial attacks~\cite{DGA-Rossow-2024}.
In counter-censorship models, explaining the results would aid in understanding the censorship strategy used~\cite{detecting-proxies-NDSS-2020, shadowsocks-China-IMC-2020,VPN-fingeprinting-2025}---such as whether the censorship technique in use depends highly on the features of the first few packets~\cite{shadowsocks-China-IMC-2020}---so as to build effective models for evading censorship. 
%From the perspective of practical deployment,   
Knowing which features contribute to the accuracy of a model~\cite{iglesias2015analysis, cost-aware-FS-IoT-2021} is useful in estimating the cost of deploying a solution. 
%Global explanations are also required to evaluate the model's bias and limitations~\cite{bad-smells-EuroSP-2024}.

\begin{cooltextbox}

\begin{itemize}
    \item  Understanding task-specific user requirement is important to determine the explainability capability we expect from a model. 
    
    \item User studies~\cite{SoK-NIDS-assessment-2023} help to gain a better understanding of  requirements from stakeholders on explainable models for specific use cases. 
\end{itemize}
\end{cooltextbox}

\subsection{Many models; convergence on the horizon}

For more than two decades, network traffic modeling progressed alongside the advancements in AI, resulting in the development of hundreds of models for \netsec tasks. We moved from simple statistical methods to ML classifiers such as Decision Trees to more complex and powerful DL models such as transformers; this evolution has been instrumental in training models with bigger datasets and larger feature space, allowing researchers to also define numerous features for various tasks.

Given that network protocols are clearly defined, and substantial  progress has been made over the past decades in developing AI-based solutions for \netsec tasks, it can be argued that the {\it fundamental} features useful for modeling are well understood. For illustration, consider a set of research works across the aforementioned \netsec tasks---anomaly detection~\cite{KitSUne-2018-NDSS, GEE-2019}, flow classification for botnet detection~\cite{FlowLens-NDSS-2021}, muti-class attack classification on IoT devices~\cite{ADEPT-2021}, multi-tab website fingerprinting~\cite{multi-tab-FP-transformer-SP2023}, flow-correlation attack on Tor networks~\cite{oh2022deepcoffea}, video fingerprinting~\cite{schuster2017beauty, sabzi2024netshaper}, dApp fingerprinting~\cite{dapp-GNN-FP-2021}, prompt inference~\cite{prompt-infer-USENIX-SEC-2024} and Tor de-anonymization attack~\cite{RECTor-2025}---the {\it raw} or {\it fundamental} packet-level features at the finest granularity used are packet-size, inter-arrival time (IAT) between packets, direction, protocol, and representations for src/dst IP addresses and port numbers. 
%\footnote{In other words, for IP address and port numbers, the differentiation is between internal vs. external node, server vs. client role, well-known vs. unknown service, etc.}. 
Features at a coarser granularity, such as flow-level features, are derived from these raw features; for example, flow volume (duration) is the sum of sizes (IAT) of packets in that flow\footnote{To be sure, there are a few exceptions. For example, DGA detection is based on DNS payloads, which however would not be available when encrypted DNS protocols are employed. Another exception is TLS feature extraction for representing session-level information. However, TLS is ubiquitous, and the transition from TLS~1.2 to TLS~1.3 has made most features unavailable, thanks to the encryption of handshakes and introduction of ECH. The estimation of certificate size that might be useful for C\&C detection~\cite{C2-TLS-RAID-2024}, could be learned by training on packet-sequences with size being one of the features.}.

\begin{cooltextbox}
State-of-the-art:
\begin{itemize}
    \item Fundamental features  from raw encrypted packets remain largely consistent across various \netsec tasks. 
%Coarser features are engineered on top of these fundamental features. 

    \item DL models today are capable of handling huge datasets and large feature space.
\end{itemize}

A natural next challenge is to go beyond having multiple models for the different tasks and build a foundation model for network traffic analysis.
\end{cooltextbox}

The concept of foundation models recently gained attention in the network community~\cite{foundation-model-thinking-HotNets-2022, ET-BERT-2022, Ptu-pre-trained-nw-traffic-ICNP-2024,llm-cybersec-2024, netfound-2025}. Once a foundation model has been built, it can be applied to different \netsec tasks, say, by fine-tuning with task-specific data. We argue that, for a foundation network model to be truly effective, it must possess several key properties that enhance its practical utility:

\begin{itemize}
    \item Different network environments---consumer, enterprise, datacenter, and telco (5G network)---have different capacities and characteristics, leading to distinct methods of traffic capture. Cost of extracting and storing traffic features is a primary constraint, influenced by the speed and scale of the network---1~Gbps consumer bandwidth with a few nodes compared to a telco network operating at hundreds of Gbps with millions of subscribers.
    Therefore, a flexible foundation model is necessary to make inferences on traffic captured at different scales represented at multiple levels: i)~raw packet features, ii)~flows or 5-tuple packet aggregates, and iii)~sessions or flow aggregates~\cite{UniNet-2025}. This flexibility enables the model to be deployed across networks with diverse data rates and capabilities. For instance, a home consumer network might have the capability to extract raw features from all packets, thereby deriving flow and session features, and thus making all {\em views} available. In contrast, a telco network, handling much higher volumes of traffic, might only collect aggregate features, such as IP-based session-level data.

    \item Following the unsupervised pretraining process in NLP~\cite{BERT-2018} (e.g., using masked language modeling), foundation network model could  be trained in a self-supervised way. However, fine-tuning to specific use cases requires labeled data. Labeling flows and packets is not only labor-intensive but also prone to errors. Therefore, a foundational model should be designed to handle missing labels during fine-tuning for specific tasks, enabling it to adapt effectively even when complete annotations are not available.

    \item As multiple foundation models are likely to emerge out of research, each potentially trained on different datasets---some openly available (e.g., MAWI~\cite{MAWI-dataset} and CAIDA~\cite{caida-2018-2019} and others private~\cite{ET-BERT-2022, Ptu-pre-trained-nw-traffic-ICNP-2024}--it becomes important to explore methods for combining these models effectively for better cost-effective utilization. Simultaneously, we must address the challenge of minimizing the risk of sensitive information leakage when integrating these models, ensuring that the benefits of collaboration are realized while maintaining data privacy and security.

    \item Explainability is an active and challenging issue in foundation models today~\cite{zhao2024explainability}. When designing a foundation model for \netsec tasks, it is crucial to prioritize explainability. The challenge lies in utilizing the foundation model for various downstream tasks while also providing clear interpretations of its decisions.
    
\end{itemize}

\begin{cooltextbox}
Key properties of a foundation model for network traffic:
\begin{itemize}
    \item Representations capturing different traffic characteristics at different scales and views.

    \item Capability to handle missing and noisy labels.

    \item Collaboration and openness, while ensuring user privacy is protected, for the effective utilization of multiple foundation models.

    \item Explainability for the downstream tasks.
\end{itemize}

\end{cooltextbox}

\section{Conclusion}
In this work, we highlighted the advancements in traffic modeling for \netsec tasks, with the aim of understanding and identifying the challenges that lie ahead. We hope this discussion serves as a catalyst for defining relevant research problems and generating new ideas, ultimately enhancing the likelihood of developing deployable and effective models for securing networks and users. \\

\section{Acknowledgment}
%\noindent \textbf{Acknowledgment}: 

This research/project is supported by the National Research Foundation, Singapore, and the Cyber Security Agency of Singapore under the National Cybersecurity R\&D Programme and the CyberSG R\&D Programme Office (Award CRPO-GC2-ASTAR-001). 
Any opinions, findings, conclusions, or recommendations expressed in these materials are those of the author(s) and do not reflect the views of the National Research Foundation, Singapore, the Cyber Security Agency of Singapore, or the CyberSG R\&D Programme Office.

\bibliography{references}

@article{ADEPT-2021,
  title={{ADEPT: Detection and Identification of Correlated Attack Stages in IoT Networks}},
  author={Sudheera, Kalupahana Liyanage Kushan and Divakaran, Dinil Mon and Singh, Rhishi Pratap and Gurusamy, Mohan},
  journal={IEEE Internet Things Journal},
  year={2021},
}

@article{evidence-gathering-2017,
title = {Evidence gathering for network security and forensics},
journal = {{Digital Investigation}},
year = {2017},
note = {DFRWS},
author = {Dinil Mon Divakaran and Kar Wai Fok and Ido Nevat and Vrizlynn L.L. Thing},
}

@INPROCEEDINGS{GEE-2019,
  author={Nguyen, Quoc Phong and Lim, Kar Wai and Divakaran, Dinil Mon and Low, Kian Hsiang and Chan, Mun Choon},
  title={{GEE: A Gradient-based Explainable Variational Autoencoder for Network Anomaly Detection}}, 
  booktitle={Proc. IEEE CNS},
  year={2019}
}

@article{PCA-Diot-2004,
    author = {Lakhina, Anukool and Crovella, Mark and Diot, Christophe},
    title = {{Diagnosing Network-Wide Traffic Anomalies}},
    year = {2004},
    volume = {34},
    number = {4},
    journal = {ACM SIGCOMM Comput. Commun. Rev.},
    month = Aug,
    pages = {219–230},
    numpages = {12},
}

@article{Levi-2024,
  title={{DNS-over-QUIC and HTTP/3 in the Era of Transformers: The New Internet Privacy Battle}},
  author={Csikor, Levente and Lian, Ziyue and Zhang, Haoran and Lakshmanan, Nitya and Divakaran, Dinil Mon},
  journal={IEEE Communications Magazine},
  year={2025}
}

@ARTICLE{NADA-2018,
    author={I. Nevat and D. M. Divakaran and S. G. Nagarajan and P. Zhang and L. Su and L. L. Ko and V. L. L. Thing},
    title={{Anomaly Detection and Attribution in Networks With Temporally Correlated Traffic}},
    journal={IEEE/ACM Trans. Netw.},
    year={2018},
    volume={26},
    number={1},
    pages={131-144},
}

@inproceedings{brauckhoff2009applying,
  title={{Applying PCA for traffic anomaly detection: Problems and solutions}},
  author={Brauckhoff, Daniela and Salamatian, Kave and May, Martin},
  booktitle={IEEE INFOCOM},
  pages={2866--2870},
  year={2009},
}

@INPROCEEDINGS{adversarially-learned-AD-ICDM-2018,
  author={Zenati, Houssam and Romain, Manon and Foo, Chuan-Sheng and Lecouat, Bruno and Chandrasekhar, Vijay},
  booktitle={IEEE ICDM}, 
  title={{Adversarially Learned Anomaly Detection}}, 
  year={2018},
}

@inproceedings{Lumos-hidden-IoT-USENIX-SEC-2022,
author = {Rahul Anand Sharma and Elahe Soltanaghaei and Anthony Rowe and Vyas Sekar},
title = {{Lumos: Identifying and Localizing Diverse Hidden {IoT} Devices in an Unfamiliar Environment}},
booktitle = {31st USENIX Security Symposium},
year = {2022},
month = aug
}

@inproceedings{Tor-malware-CCS-2022,
  title={{Exposing the rat in the tunnel: Using traffic analysis for Tor-based malware detection}},
  author={Dodia, Priyanka and AlSabah, Mashael and Alrawi, Omar and Wang, Tao},
  booktitle={Proc. CCS},
  year={2022}
}

@INPROCEEDINGS{botnet-S&P-2020,
    author={Jan, Steve T.K. and Hao, Qingying and Hu, Tianrui and Pu, Jiameng and Oswal, Sonal and Wang, Gang and Viswanath, Bimal},
    booktitle={Proc. IEEE S\&P}, 
    title={{Throwing Darts in the Dark? Detecting Bots with Limited Data using Neural Data Augmentation}}, 
    year={2020},
    pages={1190-1206},
}

@INPROCEEDINGS{WU-ZEST,
  author={Wu, Binghui and Gysel, Philipp and Divakaran, Dinil Mon and Gurusamy, Mohan},
  booktitle={IEEE NOMS}, 
  title={{ZEST: Attention-based Zero-Shot Learning for Unseen IoT Device Classification}}, 
  year={2024},
  pages={1-9}}

@inproceedings{IoT-biLSTM-2020,
  title={{Your smart home can't keep a secret: Towards automated fingerprinting of IoT traffic}},
  author={Dong, Shuaike and Li, Zhou and Tang, Di and Chen, Jiongyi and Sun, Menghan and Zhang, Kehuan},
  booktitle={Proc. ACM AisaCCS},
  year={2020}
}

@inproceedings{KitSUne-2018-NDSS,
  author       = {Yisroel Mirsky and
                  Tomer Doitshman and
                  Yuval Elovici and
                  Asaf Shabtai},
  title        = {{Kitsune: An Ensemble of Autoencoders for Online Network Intrusion Detection}},
  booktitle    = {Proc. NDSS},
  year         = {2018}
}

@misc{netreplica-2025,
      title={{NETREPLICA: Toward a Programmable Substrate for Last-Mile Data Generation}}, 
      author={Jaber Daneshamooz and Satyandra Guthula and Jessica Nguyen and William Chen and Sanjay Chandrasekaran and Ankit Gupta and Arpit Gupta and Walter Willinger},
      year={2025},
      eprint={2507.13476},
      archivePrefix={arXiv},
      url={https://arxiv.org/abs/2507.13476}, 
}

@ARTICLE{llm-cybersec-2024,
author={Divakaran, Dinil Mon and Peddinti, Sai Teja},
journal={ IEEE Security \& Privacy },
title={{Large Language Models for Cybersecurity: New Opportunities}},
year={2025},
volume={23},
number={05},
pages={38-45},
doi={10.1109/MSEC.2024.3504512},
month=sep}

@ARTICLE{hybrid-inetwork-classification-2024,
  author={Zheng, Changgang and Xiong, Zhaoqi and Bui, Thanh T. and Kaupmees, Siim and Bensoussane, Riyad and Bernabeu, Antoine and Vargaftik, Shay and Ben-Itzhak, Yaniv and Zilberman, Noa},
  journal={IEEE/ACM Transactions on Networking}, 
  title={{IIsy: Hybrid In-Network Classification Using Programmable Switches}}, 
  year={2024},
  volume={32},
  number={3},
  pages={2555-2570},
  doi={10.1109/TNET.2024.3364757}}

@inproceedings{Unsenix2023-IDP-programmable-switch-flow,
  title={{An Efficient Design of Intelligent Network Data Plane}},
  author={Zhou, Guangmeng and Liu, Zhuotao and Fu, Chuanpu and Li, Qi and Xu, Ke},
  booktitle={32nd USENIX Security Symposium },
  pages={6203--6220},
  year={2023}
}

@inproceedings{in-network-dumb-switch-hotnets-2017,
author = {Sapio, Amedeo and Abdelaziz, Ibrahim and Aldilaijan, Abdulla and Canini, Marco and Kalnis, Panos},
title = {{In-Network Computation is a Dumb Idea Whose Time Has Come}},
year = {2017},
booktitle = {Proc. HotNets}
}

@article{p4-survey-2021,
author = {Michel, Oliver and Bifulco, Roberto and R\'{e}tv\'{a}ri, G\'{a}bor and Schmid, Stefan},
journal = {ACM Comput. Surv.},
title = {{The Programmable Data Plane: Abstractions, Architectures, Algorithms, and Applications}},
year = {2021},
number = {4},
issn = {0360-0300},
url = {https://doi.org/10.1145/3447868},
doi = {10.1145/3447868},
month = may,
articleno = {82},
numpages = {36},
}

@article{offloadingML-ACM-Surveys-2023,
  title={{Offloading machine learning to programmable data planes: A systematic survey}},
  author={Parizotto, Ricardo and Coelho, Bruno Loureiro and Nunes, Diego Cardoso and Haque, Israat and Schaeffer-Filho, Alberto},
  journal={ACM Computing Surveys},
  volume={56},
  number={1},
  pages={1--34},
  year={2023},
}

@ARTICLE{UniNet-2025,
  author={Wu, Binghui and Divakaran, Dinil Mon and Gurusamy, Mohan},
  journal={IEEE Trans. on Cognitive Communications and Networking}, 
  title={{UniNet: A Unified Multi-Granular Traffic Modeling Framework for Network Security}}, 
  year={2025},
}

@article{RECTor-2025,
    author = {Binghui Wu and Dinil Divakaran and Levente Csikor and Mohan Gurusamy},
    title = {{RECTor: Robust and Efficient Correlation Attack on Tor}},
    journal = {IEEE Communications Magazine},
    year = 2025
}

@inproceedings{zhang2020poseidon,
  title={{Poseidon: Mitigating volumetric DDoS attacks with programmable switches}},
  author={Zhang, Menghao and Li, Guanyu and Wang, Shicheng and Liu, Chang and Chen, Ang and Hu, Hongxin and Gu, Guofei and Li, Qianqian and Xu, Mingwei and Wu, Jianping},
  booktitle={Proc. NDSS},
  year={2020}
}

@inproceedings{anderson2017machine,
  title={Machine learning for encrypted malware traffic classification: accounting for noisy labels and non-stationarity},
  author={Anderson, Blake and McGrew, David},
  booktitle={Proc. ACM KDD},
  year={2017}
}

@inproceedings{ET-BERT-2022,
  title={{ET-BERT: A Contextualized Datagram Representation with
Pre-training Transformers for Encrypted Traffic Classification}},
  author={Lin, Xinjie and Xiong, Gang and Gou, Gaopeng and Li, Zhen and Shi, Junzheng and Yu, Jing},
  booktitle={Proc. ACM Web Conference},
  year={2022}
}

@INPROCEEDINGS{Ptu-pre-trained-nw-traffic-ICNP-2024,
  author={Peng, Lingfeng and Xie, Xiaohui and Huang, Sijiang and Wang, Ziyi and Cui, Yong},
  booktitle={Proc. ICNP}, 
  title={{Ptu: Pre-Trained Model for Network Traffic Understanding}}, 
  year={2024},
}

@inproceedings{BERT-2018,
  author       = {Devlin, Jacob and Chang, Ming Wei and Lee, Kenton and Toutanova, Kristina},
  title        = {{BERT: Pre-training of Deep Bidirectional Transformers for Language Understanding}},
  booktitle    = {Proc. NAACL-HLT},
  year         = {2019},
}

@INPROCEEDINGS{bad-smells-EuroSP-2024,
  author={Flood, Robert and Engelen, Gints and Aspinall, David and Desmet, Lieven},
  booktitle={IEEE EuroS\&P}, 
  title={{Bad Design Smells in Benchmark NIDS Datasets}}, 
  year={2024},
}

@article{adversarial-NIDS-2022,
author = {Sheatsley, Ryan and Papernot, Nicolas and Weisman, Michael J. and Verma, Gunjan and McDaniel, Patrick},
title = {{Adversarial examples for network intrusion detection systems}},
year = {2022},
volume = {30},
number = {5},
journal = {J. Comput. Secur.},
month = jan,
pages = {727–752},
}

@techreport{IPFIX,
  author = {B. Claise and B. Trammell and P. Aitken},
  title = {Specification of the {IP} Flow Information Export ({IPFIX}) Protocol for the Exchange of Flow Information},
  howpublished = {Internet Requests for Comments},
  type = {STD},
  number = {77},
  year = {2013},
  publisher = {RFC Editor},
  institution = {RFC Editor},
  url = {http://www.rfc-editor.org/rfc/rfc7011.txt},
}

@INPROCEEDINGS{GAN-WFP-defense-SP-2022,
  author={Gong, Jiajun and Zhang, Wuqi and Zhang, Charles and Wang, Tao},
  booktitle={Proc. IEEE S\&P}, 
  title={{Surakav: Generating Realistic Traces for a Strong Website Fingerprinting Defense}}, 
  year={2022},
}

@inproceedings{wu2018wasserstein,
  title={{Wasserstein divergence for GANs}},
  author={Wu, Jiqing and Huang, Zhiwu and Thoma, Janine and Acharya, Dinesh and Van Gool, Luc},
  booktitle={Proc. ECCV},
  year={2018}
}

@article{data-explainable-WFP-PETS-2023,
  title={{Data-explainable website fingerprinting with network simulation}},
  author={Jansen, Rob and Wails, Ryan},
  journal={Proc. PETS},
  year={2023}
}

@inproceedings{retrace-Tor-2024,
author = {Jansen, Rob and Wails, Ryan and Johnson, Aaron},
title = {{Repositioning Real-World Website Fingerprinting on Tor}},
year = {2024},
booktitle = {Proc. ACM Workshop on Privacy in the Electronic Society},
}

@article{genuineTorTraces-2024,
  title={{A Measurement of Genuine Tor Traces for Realistic Website Fingerprinting}},
  author={Jansen, Rob and Wails, Ryan and Johnson, Aaron},
  journal={arXiv preprint arXiv:2404.07892},
  year={2024}
}

@inproceedings{siby2023PET,
title = {{Evaluating practical QUIC website fingerprinting defenses for the masses}},
author = "Sandra Siby and Ludovic Barman and Christopher Wood and Marwan Fayed and Nick Sullivan and Carmela Troncoso",
year = "2023",
booktitle = "Proc. PETS",
}

@inproceedings {smith2022QCSD,
author = {Jean-Pierre Smith and Luca Dolfi and Prateek Mittal and Adrian Perrig},
title = {{QCSD}: A {QUIC} {Client-Side} {Website-Fingerprinting} Defence Framework},
booktitle = {USENIX Security Symposium},
year = {2022},
}

@inproceedings {gong2020FRONTGLUE,
author = {Jiajun Gong and Tao Wang},
title = {{Zero-delay Lightweight Defenses against Website Fingerprinting}},
booktitle = {USENIX Security Symposium},
year = {2020},
}

@inproceedings{MIL-WFP-USENIX-2024,
  title={{Stop, don't click here anymore: boosting website fingerprinting by considering sets of subpages}},
  author={Mitseva, Asya and Panchenko, Andriy},
  booktitle={Proc. USENIX Security Symposium},
  year={2024}
}

@inproceedings{shenoi2023ipet,
  title={{iPET: Privacy Enhancing Traffic Perturbations for Secure IoT Communications}},
  author={Shenoi, Akshaye and Vairam, Prasanna Karthik and Sabharwal, Kanav and Li, Jialin and Divakaran, Dinil Mon},
  booktitle={Proc. PETS},
  year={2023}
}

@inproceedings{poursaeed2018generative,
  title={Generative adversarial perturbations},
  author={Poursaeed, Omid and Katsman, Isay and Gao, Bicheng and Belongie, Serge},
  booktitle={Proc. CVPR},
  year={2018}
}

@inproceedings{oh2022deepcoffea,
  title={{DeepCoFFEA: Improved flow correlation attacks on Tor via metric learning and amplification}},
  author={Oh, Se Eun and Yang, Taiji and Mathews, Nate and Holland, James K and Rahman, Mohammad Saidur and Hopper, Nicholas and Wright, Matthew},
  booktitle={Proc. IEEE S\&P},
  year={2022},
}

@inproceedings{nasr2018deepcorr,
  title={{DeepCorr: Strong flow correlation attacks on tor using deep learning}},
  author={Nasr, Milad and Bahramali, Alireza and Houmansadr, Amir},
  booktitle={Proc. ACM CCS},
  pages={1962--1976},
  year={2018}
}

@inproceedings{dingledine2004tor,
  title={{Tor: The second-generation onion router}},
  author={Dingledine, Roger and Mathewson, Nick and Syverson, Paul F and others},
  booktitle={Proc. USENIX security symposium},
  volume={4},
  pages={303--320},
  year={2004}
}

@article{survey-encrypted-DNS-mw-2022,
author = {Lyu, Minzhao and Gharakheili, Hassan Habibi and Sivaraman, Vijay},
title = {{A Survey on DNS Encryption: Current Development, Malware Misuse, and Inference Techniques}},
year = {2022},
volume = {55},
number = {8},
issn = {0360-0300},
url = {https://doi.org/10.1145/3547331},
doi = {10.1145/3547331},
journal = {ACM Comput. Surv.},
month = dec
}

@inproceedings{real-time-DNS-exfiltration-NDSS-2024,
  title={{Information based heavy hitters for real-time DNS data exfiltration detection}},
  author={Ozery, Yarin and Nadler, Asaf and Shabtai, Asaf},
  booktitle={Proc. NDSS},
  pages={1--15},
  year={2024}
}

@inproceedings{transformer-WFP-CCS-2023,
author = {Jin, Zhaoxin and Lu, Tianbo and Luo, Shuang and Shang, Jiaze},
title = {{Transformer-based Model for Multi-tab Website Fingerprinting Attack}},
year = {2023},
booktitle = {Proc. ACM CCS},
}

@inproceedings{Google-transparency-mail,
booktitle = {{Google Transparency Report}},
title = {{Email encryption in transit}},
author = {Google},
year = {2025},
url = {https://transparencyreport.google.com/safer-email/overview}
}

@inproceedings{Google-transparency-HTTP,
booktitle = {{Google Transparency Report}}, 
title = {{HTTPS encryption on the web}},
author = {Google},
year = {2025},
url = {https://transparencyreport.google.com/https/overview}
}

@inproceedings{Usenix-DDos-programme-switch-2021,
  title={{Jaqen: A High-Performance Switch-Native approach for detecting and mitigating volumetric DDoS attacks with programmable switches}},
  author={Liu, Zaoxing and Namkung, Hun and Nikolaidis, Georgios and Lee, Jeongkeun and Kim, Changhoon and Jin, Xin and Braverman, Vladimir and Yu, Minlan and Sekar, Vyas},
  booktitle={Proc. USENIX Security Symposium},
  year={2021}
}

@ARTICLE{Marina-ML-monitoring-Terabit-2024,
  author={Seufert, Michael and Dietz, Katharina and Wehner, Nikolas and Geißler, Stefan and Schüler, Joshua and Wolz, Manuel and Hotho, Andreas and Casas, Pedro and Hoßfeld, Tobias and Feldmann, Anja},
  journal={IEEE Trans. on Network and Service Management}, 
  title={{Marina: Realizing ML-Driven Real-Time Network Traffic Monitoring at Terabit Scale}}, 
  year={2024},
  volume={21},
  number={3},
  pages={2773-2790},
  doi={10.1109/TNSM.2024.3382393}}

@article{sampling-elephants-2003,
  title={{New directions in traffic measurement and accounting: Focusing on the elephants, ignoring the mice}},
  author={Estan, Cristian and Varghese, George},
  journal={ACM Trans. on Computer Systems (TOCS)},
  volume={21},
  number={3},
  pages={270--313},
  year={2003}
}

@inproceedings{2019-do-switchdream,
  title={{Do switches dream of machine learning? Toward in-network classification}},
  author={Xiong, Zhaoqi and Zilberman, Noa},
  booktitle={Proc. ACM HotNets},
  pages={25--33},
  year={2019}
}

@inproceedings{Usenix-2023-website-fingerprinting,
  title={{Subverting website fingerprinting defenses with robust traffic representation}},
  author={Shen, Meng and Ji, Kexin and Gao, Zhenbo and Li, Qi and Zhu, Liehuang and Xu, Ke},
  booktitle={Proc. USENIX Security Symposium},
  year={2023}
}

@inproceedings{kohls2019lost,
  title={{Lost traffic encryption: fingerprinting LTE/4G traffic on layer two}},
  author={Kohls, Katharina and Rupprecht, David and Holz, Thorsten and P{\"o}pper, Christina},
  booktitle={Proceedings of the 12th ACM Conference on Security and Privacy in Wireless and Mobile Networks (WiSec)},
  pages={249--260},
  year={2019}
}

@inproceedings{vaswani2017attention,
  title={{Attention Is All You Need}},
  author={Vaswani, Ashish and Shazeer, Noam and Parmar, Niki and Uszkoreit, Jakob and Jones, Llion and Gomez, Aidan N and Kaiser, {\L}ukasz and Polosukhin, Illia},
  booktitle={Proc. NIPS},
  year={2017}
}

@article{webpage_quic,
  author       = {Jean{-}Pierre Smith and
                  Prateek Mittal and
                  Adrian Perrig},
  title        = {Website Fingerprinting in the Age of {QUIC}},
  journal      = {Proc. Priv. Enhancing Technol.},
  volume       = {2021},
  number       = {2},
  pages        = {48--69},
  year         = {2021}
}

@inproceedings{unknown-app-classification-2020,
  title={{Autonomous Unknown-Application Filtering and Labeling for DL-based Traffic Classifier Update}},
  author={Zhang, Jielun and Li, Fuhao and Ye, Feng and Wu, Hongyu},
  booktitle={Proc. IEEE INFOCOM},
  pages={397--405},
  year={2020}
}

@inproceedings{AD-robust-AE-2017,
author = {Zhou, Chong and Paffenroth, Randy C.},
title = {Anomaly Detection with Robust Deep Autoencoders},
year = {2017},
url = {https://doi.org/10.1145/3097983.3098052},
doi = {10.1145/3097983.3098052},
booktitle = {Proc. ACM KDD}
}

@misc{T5,
      title={Exploring the Limits of Transfer Learning with a Unified Text-to-Text Transformer}, 
      author={Colin Raffel and Noam Shazeer and Adam Roberts and Katherine Lee and Sharan Narang and Michael Matena and Yanqi Zhou and Wei Li and Peter J. Liu},
      year={2020},
      eprint={1910.10683},
      archivePrefix={arXiv},
      primaryClass={cs.LG}
}

@article{zhao2024explainability,
  title={Explainability for large language models: A survey},
  author={Zhao, Haiyan and Chen, Hanjie and Yang, Fan and Liu, Ninghao and Deng, Huiqi and Cai, Hengyi and Wang, Shuaiqiang and Yin, Dawei and Du, Mengnan},
  journal={ACM Transactions on Intelligent Systems and Technology},
  volume={15},
  number={2},
  pages={1--38},
  year={2024},
}

@inproceedings{traffic-classification-KDD-2023,
author = {Fauvel, Kevin and Chen, Fuxing and Rossi, Dario},
title = {{A Lightweight, Efficient and Explainable-by-Design Convolutional Neural Network for Internet Traffic Classification}},
year = {2023},
url = {https://doi.org/10.1145/3580305.3599762},
doi = {10.1145/3580305.3599762},
booktitle = {Proc. ACM KDD}
}

@article{pfp-PETS-2019,
  title={p-FP: Extraction, classification, and prediction of website fingerprints with deep learning},
  author={Oh, Se Eun and Sunkam, Saikrishna and Hopper, Nicholas},
  journal={Proc. PETS},
  year={2019}
}

@ARTICLE{DEFT-2019,
  author={Thangavelu, Vijayanand and Divakaran, Dinil Mon and Sairam, Rishi and Bhunia, Suman Sankar and Gurusamy, Mohan},
  journal={IEEE Internet of Things Journal}, 
  title={DEFT: A Distributed IoT Fingerprinting Technique}, 
  year={2019},
}

@inproceedings{Usenix-2020-youarewhatyou,
  title={{You are what you broadcast: Identification of mobile and IoT devices from (public) WiFi}},
  author={Yu, Lingjing and Luo, Bo and Ma, Jun and Zhou, Zhaoyu and Liu, Qingyun},
  booktitle={Proc. USENIX Security Symposium},
  year={2020}
}

@article{varCNN-WFP-2019,
  author       = {Sanjit Bhat and
                  David Lu and
                  Albert Kwon and
                  Srinivas Devadas},
  title        = {{Var-CNN: {A} Data-Efficient Website Fingerprinting Attack Based on
                  Deep Learning}},
  journal      = {Priv. Enhancing Technol.},
  volume       = {2019},
  number       = {4},
  pages        = {292--310},
  year         = {2019},
}

@inproceedings{Usenix-2022-onlinewebsitefingerprinting,
  title={{Online website fingerprinting: Evaluating website fingerprinting attacks on tor in the real world}},
  author={Cherubin, Giovanni and Jansen, Rob and Troncoso, Carmela},
  booktitle={Proc. USENIX Security Symposium},
  year={2022}
}

@inproceedings{FlowLens-NDSS-2021,
  title={{FlowLens: Enabling Efficient Flow Classification for ML-based Network Security Applications.}},
  author={Barradas, Diogo and Santos, Nuno and Rodrigues, Lu{\'\i}s and Signorello, Salvatore and Ramos, Fernando MV and Madeira, Andr{\'e}},
  booktitle={Proc. NDSS},
  year={2021}
}

@inproceedings{2017-profileiot,
  title={{ProfilIoT: A machine learning approach for IoT device identification based on network traffic analysis}},
  author={Meidan, Yair and Bohadana, Michael and Shabtai, Asaf and Guarnizo, Juan David and Ochoa, Mart{\'\i}n and Tippenhauer, Nils Ole and Elovici, Yuval},
  booktitle={Proc. of the Symposium on Applied Computing},
  pages={506--509},
  year={2017}
}

@inproceedings{2019-DIoT,
  title={D{\"I}oT: A federated self-learning anomaly detection system for IoT},
  author={Nguyen, Thien Duc and Marchal, Samuel and Miettinen, Markus and Fereidooni, Hossein and Asokan, Nadarajah and Sadeghi, Ahmad-Reza},
  booktitle={Proc. ICDCS},
  year={2019},
}

@inproceedings{global-explanations-AD-CCS-2024,
author = {Han, Dongqi and Wang, Zhiliang and Feng, Ruitao and Jin, Minghui and Chen, Wenqi and Wang, Kai and Wang, Su and Yang, Jiahai and Shi, Xingang and Yin, Xia and Liu, Yang},
title = {Rules Refine the Riddle: Global Explanation for Deep Learning-Based Anomaly Detection in Security Applications},
year = {2024},
booktitle = {Proc. ACM CCS}
}

@inproceedings {xNIDS-USENIX-SEC-2023,
author = {Feng Wei and Hongda Li and Ziming Zhao and Hongxin Hu},
title = {{{xNIDS}: Explaining Deep Learning-based Network Intrusion Detection Systems for Active Intrusion Responses}},
booktitle = {Proc. USENIX Security Symposium},
year = {2023},
}

@inproceedings{NDSS-2018-wfp-lstm,
  author       = {Vera Rimmer and
                  Davy Preuveneers and
                  Marc Juarez and
                  Tom van Goethem and
                  Wouter Joosen},
  title        = {{Automated Website Fingerprinting through Deep Learning}},
  booktitle    = {Proc. NDSS},
  year         = {2018},
}

@inproceedings{multi-tab-FP-transformer-SP2023,
  title={Robust multi-tab website fingerprinting attacks in the wild},
  author={Deng, Xinhao and Yin, Qilei and Liu, Zhuotao and Zhao, Xiyuan and Li, Qi and Xu, Mingwei and Xu, Ke and Wu, Jianping},
  booktitle={IEEE S\&P},
  pages={1005--1022},
  year={2023},
}

@article{XAI-malware-2024,
  title={{A Comprehensive Analysis of Explainable AI for Malware Hunting}},
  author={Saqib, Mohd and Mahdavifar, Samaneh and Fung, Benjamin CM and Charland, Philippe},
  journal={ACM Computing Surveys},
  volume={56},
  number={12},
  pages={1--40},
  year={2024},
}

@inproceedings{lee-LLM-phishing-2024,
      title={{Multimodal Large Language Models for Phishing Webpage Detection and Identification}}, 
      author={Jehyun Lee and Peiyuan Lim and Bryan Hooi and Dinil Mon Divakaran},
      year={2024},
      booktitle={Proc. Symposium on Electronic Crime Research (eCrime)}
}

@inproceedings {prov-ninja-2023,
author = {Kunal Mukherjee and Joshua Wiedemeier and Tianhao Wang and James Wei and Feng Chen and Muhyun Kim and Murat Kantarcioglu and Kangkook Jee},
title = {Evading {Provenance-Based} {ML} Detectors with Adversarial System Actions},
booktitle = {Proc. USENIX Security Symposium},
year = {2023}
}

@inproceedings{MAWI-dataset,
author = {Kenjiro Cho and Koushirou Mitsuya and Akira Kato},
title = {{Traffic Data Repository at the {WIDE} Project}},
booktitle = {Proc. USENIX ATC},
year = {2000},
month = jun
}

@misc{caida-2018-2019,
  title={{The CAIDA Anonymized Internet Traces Dataset (April 2008 - January 2019)}},
  author={{The CAIDA UCSD}},
  year = 2025,
  url = {{https://www.caida.org/catalog/datasets/passive\_dataset/}}
}

@inproceedings {prompt-infer-USENIX-SEC-2024,
author = {Roy Weiss and Daniel Ayzenshteyn and Yisroel Mirsky},
title = {{What Was Your Prompt? A Remote Keylogging Attack on {AI} Assistants}},
booktitle = {Proc. USENIX Security Symposium},
year = {2024},
}

@inproceedings{shadowsocks-China-IMC-2020,
author = {Alice and Bob and Carol and Beznazwy, Jan and Houmansadr, Amir},
title = {{How China Detects and Blocks Shadowsocks}},
year = {2020},
booktitle = {Proc. IMC}
}

@article{VPN-fingeprinting-2025,
author = {Xue, Diwen and Ramesh, Reethika and Jain, Arham and Kallitsis, Michaelis and Halderman, J. Alex and Crandall, Jedidiah R. and Ensafi, Roya},
title = {{OpenVPN is Open to VPN Fingerprinting}},
year = {2024},
journal = {Commun. ACM},
month = dec,
pages = {79–87},
numpages = {9}
}

@inproceedings{detecting-proxies-NDSS-2020,
  title={{Detecting Probe-resistant Proxies}},
  author={Frolov, Sergey and Wampler, Jack and Wustrow, Eric},
  booktitle={Proc. NDSS},
  year={2020}
}

@inproceedings{cellular-video-2022,
  title={{Watching the watchers: Practical video identification attack in LTE networks}},
  author={Bae, Sangwook and Son, Mincheol and Kim, Dongkwan and Park, CheolJun and Lee, Jiho and Son, Sooel and Kim, Yongdae},
  booktitle={Proc. USENIX Security Symposium},
  year={2022}
}

@inproceedings{schuster2017beauty,
  title={{Beauty and the burst: Remote identification of encrypted video streams}},
  author={Schuster, Roei and Shmatikov, Vitaly and Tromer, Eran},
  booktitle={Proc. USENIX Security Symposium},
  year={2017}
}

@inproceedings{gast2024snailload,
  title={{$\{$SnailLoad$\}$: Exploiting Remote Network Latency Measurements without $\{$JavaScript$\}$}},
  author={Gast, Stefan and Czerny, Roland and Juffinger, Jonas and Rauscher, Fabian and Franza, Simone and Gruss, Daniel},
  booktitle={Proc. USENIX Security Symposium},
  year={2024}
}

@inproceedings{sabzi2024netshaper,
  title={{NetShaper: A Differentially Private Network Side-Channel Mitigation System}},
  author={Sabzi, Amir and Vora, Rut and Goswami, Swati and Seltzer, Margo and L{\'e}cuyer, Mathias and Mehta, Aastha},
  booktitle={Proc. USENIX Security Symposium },
  year={2024}
}

@article{dapp-GNN-FP-2021,
  title={Accurate decentralized application identification via encrypted traffic analysis using graph neural networks},
  author={Shen, Meng and Zhang, Jinpeng and Zhu, Liehuang and Xu, Ke and Du, Xiaojiang},
  journal={IEEE Trans. on Information Forensics and Security},
  year={2021},
}

@inproceedings{foundation-model-thinking-HotNets-2022,
author = {Le, Franck and Srivatsa, Mudhakar and Ganti, Raghu and Sekar, Vyas},
title = {Rethinking data-driven networking with foundation models: challenges and opportunities},
year = {2022},
booktitle = {Proc ACM HotNets}
}

@inproceedings{DGA-Rossow-2024,
author = {Cebere, Bogdan Constantin and Flueren, Jonathan Lasse Bennet and Sebasti\'{a}n, Silvia and Plohmann, Daniel and Rossow, Christian},
title = {{Down to earth! Guidelines for DGA-based Malware Detection}},
year = {2024},
booktitle = {Proc. RAID}
}

@inproceedings{FS-Net-traffic-classifier-INFOCOM-2019,
  title={{FS-Net: A flow sequence network for encrypted traffic classification}},
  author={Liu, Chang and He, Longtao and Xiong, Gang and Cao, Zigang and Li, Zhen},
  booktitle={Proc. IEEE INFOCOM},
  year={2019},
}

@inproceedings{DDoS-eval-latency-ICNP-2024,
  title={{E-DDoS: An Evaluation System for DDoS Attack Detection}},
  author={Chi, Kaiwen and Xie, Xiaohui and Hu, Yannan and Zhao, Dongyang and Xie, Yuming and Zhang, Liang and Cui, Yong},
  booktitle={Proc. IEEE ICNP},
  year={2024},
}

@ARTICLE{cost-aware-FS-IoT-2021,
  author={Chakraborty, Biswadeep and Divakaran, Dinil Mon and Nevat, Ido and Peters, Gareth W. and Gurusamy, Mohan},
  journal={IEEE Internet of Things Journal}, 
  title={{Cost-Aware Feature Selection for IoT Device Classification}}, 
  year={2021}
}

@article{iglesias2015analysis,
  title={Analysis of network traffic features for anomaly detection},
  author={Iglesias, F{\'e}lix and Zseby, Tanja},
  journal={Machine Learning},
  pages={59--84},
  year={2015},
}

@inproceedings{C2-TLS-RAID-2024,
  title={{Extending C2 Traffic Detection Methodologies: From TLS 1.2 to TLS 1.3-enabled Malware}},
  author={Barradas, Diogo and Novo, Carlos and Portela, Bernardo and Romeiro, Sofia and Santos, Nuno},
  booktitle={Proc. RAID},
  year={2024}
}

@inproceedings{data-synthesis-DCN-HotNets-2023,
  title={{Datacenter Network Deserves Better Traffic Models}},
  author={Huang, Sijiang and Peng, Lingfeng and Wang, Mowei and Liu, Yashe and Liu, Zhenhua and Wang, Xin and Cui, Yong},
  booktitle={Proc. ACM HotNets},
  year={2023}
}

@inproceedings{Zoom2Net-SIGCOMM-2024,
author = {Gong, Fengchen and Raghunathan, Divya and Gupta, Aarti and Apostolaki, Maria},
title = {{Zoom2Net: Constrained Network Telemetry Imputation}},
year = {2024},
booktitle = {Proc. ACM SIGCOMM Conference},
}

@inproceedings{GAN-pkt-synthesis-SIGCOMM-2022,
  title={{Practical GAN-based synthetic ip header trace generation using netshare}},
  author={Yin, Yucheng and Lin, Zinan and Jin, Minhao and Fanti, Giulia and Sekar, Vyas},
  booktitle={Proc. ACM SIGCOMM},
  year={2022}
}

@article{jiang2024netdiffusion,
  title={{Netdiffusion: Network data augmentation through protocol-constrained traffic generation}},
  author={Jiang, Xi and Liu, Shinan and Gember-Jacobson, Aaron and Bhagoji, Arjun Nitin and Schmitt, Paul and Bronzino, Francesco and Feamster, Nick},
  journal={Proc.  of the ACM on Measurement and Analysis of Computing Systems},
  volume={8},
  number={1},
  pages={1--32},
  year={2024}
}

@article{flowChronicle-2024,
author = {C\"{u}ppers, Joscha and Schoen, Adrien and Blanc, Gregory and Gimenez, Pierre-Francois},
title = {{FlowChronicle: Synthetic Network Flow Generation through Pattern Set Mining}},
year = {2024},
volume = {2},
number = {CoNEXT4},
url = {https://doi.org/10.1145/3696407},
doi = {10.1145/3696407},
journal = {Proc. ACM Netw.},
month = nov,
articleno = {26},
numpages = {20},
}

@inproceedings{Trustee-CCS-2022,
author = {Jacobs, Arthur S. and Beltiukov, Roman and Willinger, Walter and Ferreira, Ronaldo A. and Gupta, Arpit and Granville, Lisandro Z.},
title = {{AI/ML for Network Security: The Emperor has no Clothes}},
year = {2022},
url = {https://doi.org/10.1145/3548606.3560609},
doi = {10.1145/3548606.3560609},
booktitle = {Proc. ACM CCS},
}

@inproceedings{gen-nw-traffic-2022,
author = {B\"{u}hler, Tobias and Schmid, Roland and Lutz, Sandro and Vanbever, Laurent},
title = {Generating representative, live network traffic out of millions of code repositories},
year = {2022},
url = {https://doi.org/10.1145/3563766.3564084},
doi = {10.1145/3563766.3564084},
booktitle = {Proc ACM HotNets},
}

@inproceedings{khan2024harnessing,
  title={{Harnessing Public Code Repositories to Develop Production-Ready ML Artifacts for Networking}},
  author={Khan, Punnal Ismail and Guthula, Satyandra and Beltiukov, Roman and Schmid, Roland and B{\"u}hler, Tobias and Gupta, Arpit and Vanbever, Laurent and Willinger, Walter},
  booktitle={Proc. Applied Networking Research Workshop},
  pages={100--102},
  year={2024}
}

@misc{netfound-2025,
      title={{netFound: Foundation Model for Network Security}}, 
      author={Satyandra Guthula and Roman Beltiukov and Navya Battula and Wenbo Guo and Arpit Gupta and Inder Monga},
      year={2025},
      eprint={2310.17025},
      archivePrefix={arXiv},
      primaryClass={cs.NI},
      url={https://arxiv.org/abs/2310.17025}, 
}

@inproceedings{XAI-depoloyment-2020,
author = {Bhatt, Umang and Xiang, Alice and Sharma, Shubham and Weller, Adrian and Taly, Ankur and Jia, Yunhan and Ghosh, Joydeep and Puri, Ruchir and Moura, Jos\'{e} M. F. and Eckersley, Peter},
title = {Explainable machine learning in deployment},
year = {2020},
booktitle = {Proc. Conf. on Fairness, Accountability, and Transparency},
}

@inproceedings{SoK-NIDS-assessment-2023,
  title={SoK: Pragmatic assessment of machine learning for network intrusion detection},
  author={Apruzzese, Giovanni and Laskov, Pavel and Schneider, Johannes},
  booktitle={Proc. IEEE EuroS\&P},
  pages={592--614},
  year={2023},
}

% Check whether the conference requires a reproducibility checklist to be included in the paper.
% If so, you can uncomment the following line and ajust the path to include it.
% \input{../../ReproducibilityChecklist/LaTeX/ReproducibilityChecklist.tex}

\end{document}